\documentstyle[preprint, aps]{revtex}
\tightenlines

\begin{document}
\draft
\preprint{HEP/123-qed}
\title{Origin and pressure dependence of ferromagnetism in $A_{2}Mn_{2}O_{7}$ pyrochlores ($A=Y$, $In$, $Lu$, and $Tl$)\\
}
\author{M. D. N\'{u}\~{n}ez-Regueiro\\}
\address{
Grenoble High Field Magnetic Laboratory, MPI-FKF and CNRS, Bo\^{i}te Postale 166, 38042 Grenoble Cedex 9, France.\\}
\author{C. Lacroix}
\address{
Laboratoire Louis Néel, CNRS, Bo\^{i}te Postale 166, 38042 Grenoble Cedex 9, France\\}
\date{\today}
\maketitle
\begin{abstract}
Non-conventional mechanisms have been recently invoked in order to explain the ferromagnetic ground state of  $A_{2}Mn_{2}O_{7}$ pyrochlores ($A=Y$, $In$, $Lu$, and $Tl$) and the puzzling decrease of their Curie temperatures with applied pressure.
Here we show, using a perturbation expansion in the  $Mn-O$ hopping term, that both features can be understood within the superexchange model, provided that the intra-atomic oxygen interactions are properly taken into account.
An additional coupling between the $Mn$ ions mediated by the $In(5s)/Tl(6s)$ bands yields the higher $T_{C}$'s of these two compounds,
this mechanism enhancing their ferromagnetism for higher pressures.
\end{abstract}


\narrowtext

The observation of colossal magnetoresistance (CMR) in $Tl_{2-x}In_{x}Mn_{2}O_{7}$ pyrochlores \cite{shimakawa96,cheong96,subramanian96} challenges our understanding of this phenomenon with important technological applications. In fact, in contrast with their perovskite counterparts, these systems do not present mixed  $Mn^{3+}-Mn^{4+}$  valences, they have a small number of conduction carriers ($\sim10^{-3}$ per formula unit) and they do not exhibit Jahn-Teller effect or anomalous spin diffusion, denying both, the usual double exchange and the polaronic mechanisms invoked in that case. Furthermore, in spite of the similar CMR effect observed, the magnetic transition does not occur between a low temperature metal and a high temperature semiconducting state like in the perovskites, but both, the ferro and the paramagnetic phases show metallic
behaviour.\\
These compounds belong to the larger family of $A_{2}Mn_{2}O_{7}$  pyrochlores, the metallic behaviour of the  $A=Tl$ compound being the exception, since all other members are insulators. But they all show long range ferromagnetic ordering, and can be classified in two groups according to their transition temperatures \cite{shimakawa99}: $T_{C}\sim15K$ for $A=Y$ and $Lu$, while $T_{C}\sim125K$  for
$A=In$ and $Tl$.\\
Considering the $Mn^{4+}-O(I)-Mn^{4+}$ bond angle $\theta$, a ferromagnetic superexchange picture between nearest neighbour (n.n.) $Mn$ ions has been initially suggested \cite{subramanian96,shimakawa99}. In fact, following the Goodenough-Kanamori-Anderson rules \cite{khomskii97}, the $\theta\sim132^{\circ}$  of  this pyrochlore structure falls into the range in which a sign reversal of the exchange constants is expected.
Using the phenomenological approach for the angular dependence of the exchange interactions for the similar case of $Cr^{3+}-O^{2-}-Cr^{3+}$ ions \cite{motida70}, Shimakawa et al. \cite{shimakawa99} found a slightly ferromagnetic exchange $J$ interaction, in agreement with the low Curie temperatures observed for $Y_{2}Mn_{2}O_{7}$  and  $Lu_{2}Mn_{2}O_{7}$.
Furthermore, without proposing a particular mechanism, they pointed out the possible role of the hybridisation of the $In(5s)/Tl(6s)-O(2p)-Mn(3d)$ orbitals in enhancing  $T_{C}$ in these two compounds \cite{shimakawa99}.\\
On the other hand, taking the ferromagnetic ordering as an experimental fact, different theories attempted to explain the CMR effect. Scattering against ferromagnetic fluctuations, relevant in this case with a very low number of carriers \cite{majumdar98}; or simply hybridisation of the localised moments with the itinerant $Tl(6s)$ states have been invoked \cite{ventura97} as possible mechanisms for the
CMR.\\
However, recent measurements showing a decrease of the ferromagnetic $T_{C}$ with increasing applied pressure for all $A_{2}Mn_{2}O_{7}$ compounds \cite{sushko99}, have been interpreted as contradicting the original superexchange picture. The data establish that materials with lower $T_{C}$  posses a larger negative pressure shift ( $dT_{C}/dP=-3.8$, $-2.5$, $-1.6$ and  $-0.4K/GPa$, and $dlnT_{C}/dP=-24.5$, $-20.8$, $-11.2$ and  $-0.3/GPa$, for $A=Lu$, $In$, $Tl$, and  $In$, respectively).
A more exotic scenario has been suggested, with antiferromagnetic (AF) coupling between n.n. $Mn$  ions overcome by longer range ferromagnetic interactions, due to the frustration of  the former in the pyrochlore lattice \cite{sushko99}. Also the enhancement of the ferromagnetic coupling in  $Sb$-substituted  $Tl_{2}Mn_{2}O_{7}$ compounds has been discussed \cite{alonso99} as strongly supporting this new proposal. But taking into account the structural parameters of  these systems, the justification of  the necessary exchange constants
seems difficult.\\
Here we show that the observed weakening of the ferromagnetism with moderate pressures can be understood within the conventional superexchange model, provided that the intra-atomic Hund interaction $J_{H}^{Ox}$  of the oxygen has a significant value while the excitation energy  $\Delta$ of the $O(2p)$ electrons to the empty $e_{g}$  levels of the localised $Mn$ ions is small. Comparison with other oxides should indicate that these can be reasonable assumptions. The resulting exchange interaction  $J$ is ferromagnetic due to the $Mn^{4+}-O(I)-Mn^{4+}$ bond angle, and when an external pressure is applied $\Delta$  increases while the intra-atomic parameter $J_{H}^{Ox}$  remains unchanged, inducing the observed decrease of
$T_{C}$. However, an additional mechanism is still necessary to account for the observed $T_{C}$'s above $120K$.
We propose a superexchange coupling between the $Mn(t_{2g})$ states mediated by the $In(5s)/Tl(6s)-O(2p)$ bands, which also explains the increase of $T_{C}$ reported later for higher pressures \cite{n\'{u\~{n}ez99}}.
Although the relevant bands become strongly spin-polarized, the ferromagnetism and the conduction in the $Tl_{2}Mn_{2}O_{7}$ compound appear to be less coupled than in the perovskites, in agreement with
experimental results \cite{ramirez97,hwang97}.\\
We now calculate the exchange interaction between two manganese ions of the pyrochlore lattice, forming an angle $\theta$  between them through an intermediate full oxygen atom $O(I)$. Each $Mn$ ion has an inert $t_{2g}$  core with $s=3/2$  and, for simplification, let us consider just a single empty  $e_{g}$  orbital state (in fact, only the orbital directed towards $O(I)$ is relevant), onto which an $O(2p)$ electron can hope, provided that it has the same spin orientation, e.g. the Hund's rule energy cost on the $Mn$  ion  $J_{H}^{Mn}$  is taken as infinite, avoiding hops of antiparallel spins (therefore it is not necessary to include the Coulomb repulsion on the $Mn$  ions).
The $O(I)$ ions mediate the superexchange interaction, the four $2p$ electrons moving back and forth between the $O$ and the two n.n. $Mn$ orbitals within the same plane.
We call $t$ the $Mn-O(I)$  hopping matrix element, $\Delta$  the energy difference between the $O(2p)$ levels $(p_{x}, p_{y})$ and the $e_{g}$ state of each $Mn$ site $(d_{1}, d_{2})$, $J_{H}^{Ox}$  is the Hund's rule energy on the $O$ ions and $U_{p}$ the Coulomb repulsion between two holes on the same $O$  atom. These two last parameters are of crucial importance since they will be responsible for the ferromagnetic interaction for the relevant $Mn-O(I)-Mn$  angles
$\theta$ ($\theta\sim130.8^{\circ}$, $131.4^{\circ}$, $132.4^{\circ}$ and $133.4^{\circ}$ for $A=In$, $Lu$, $Y$ and $Tl$, respectively)\cite{shimakawa99}.\\
When the core spins of two n.n. $Mn$  ions $\overrightarrow{s_{1}}$  and $\overrightarrow{s_{2}}$  are ferromagnetically aligned, the possible occupations for the parallel  hopping spins (e.g. up spins) are:
$\mid d_{1}p_{x}>$, $\mid d_{1}p_{y}>$, $\mid d_{1}d_{2}>$, $\mid p_{x}p_{y}>$, $\mid p_{x}d_{2}>$ and $\mid p_{y}d_{2}>$.
The corresponding Hamiltonian reads:

\begin{equation}
H_{\uparrow\uparrow}=
\left(
  \begin{array}{cccccc}
    \Delta &      0 &                   -t_{2} & -t_{1} &      0 &     0 \\
         0 & \Delta &                    t_{1} &  t_{2} &      0 &     0 \\
    -t_{2} &   t_{1}& 2\Delta+U_{p}-J_{H}^{Ox} &     0  &  t_{2} & t_{1} \\
    -t_{1} &   t_{2}&                       0  &     0  &  t_{1} & t_{2} \\
         0 &      0 &                    t_{2} &  t_{1} & \Delta &     0 \\
         0 &      0 &                    t_{1} &  t_{2} &      0 & \Delta
\end{array}
\right)
\end{equation}
where $t_{1}\equiv t\cos(\theta/2)$  and  $t_{2}\equiv t\sin(\theta/2)$,  which we solve for $t/\Delta<1$  up to fourth order in the hopping element $t$.
In previous calculations \cite{mishra98} the $J_{H}^{Ox}$  and $U_{p}$ terms have been omitted, loosing the contribution that can give rise to the ferromagnetic interaction.
We obtain:

\begin{eqnarray}
E_{\uparrow\uparrow}=-\frac{2t^{2}}{\Delta}+\frac{2t^{4}[1+cos^{2}(\theta)]}{\Delta^{2}(\Delta+\frac{U_{p}}{2}-\frac{J_{H}^{Ox}}{2})}+{}\nonumber\\
+\frac{2t^{4}(U_{p}-J_{H}^{Ox})}{\Delta^{3}(\Delta+\frac{U_{p}}{2}-\frac{J_{H}^{Ox}}{2})}
\end{eqnarray}

Instead, when the two n.n. core $Mn$ ions are antiparallel (e.g. $\overrightarrow{s_{1}}=up$ and $\overrightarrow{s_{2}}=down$), there are three spin up and three spin down possible occupations for the four oxygen hopping electrons \cite{mishra98}:
$\mid~d_{1}p_{x},d_{2}p_{x}>$,  $\mid~d_{1}p_{x},d_{2}p_{y}>$,  $\mid~d_{1}p_{x},p_{x}p_{y}>$,  $\mid~d_{1}p_{y},d_{2}p_{x}>$,  $\mid~d_{1}p_{x},d_{2}p_{y}>$,  $\mid~d_{1}p_{y},p_{x}p_{y}>$,  $\mid~p_{x}p_{y},d_{2}p_{x}>$, $\mid~p_{x}p_{y},d_{2}p_{y}>$  and  $\mid~p_{x}p_{y},p_{x}p_{y}>$,
where now the two first index correspond to the spins up and those after the comma to the spins down occupations, respectively.
They yield the following Hamiltonian:

\widetext
\begin{equation}
H_{\uparrow\downarrow}=
\left(
\begin{array}{ccccccccc}
    2\Delta+U_{p} &             0 & -t_{1} &             0 &             0 &      0 & -t_{1} &      0 &     0\\
                0 & 2\Delta+U_{p} & -t_{2} &    J_{H}^{Ox} &             0 &      0 &      0 & -t_{1} &     0\\
           -t_{1} &        -t_{2} & \Delta &             0 &             0 &      0 &      0 &      0 & -t_{1}\\
                0 &    J_{H}^{Ox} &      0 & 2\Delta+U_{p} &             0 & -t_{1} &  t_{2} &      0 &     0\\
                0 &             0 &      0 &             0 & 2\Delta+U_{p} & -t_{2} &      0 &  t_{2} &     0\\
                0 &             0 &      0 &        -t_{1} &       -t_{ 2} & \Delta &      0 &      0 &  t_{2}\\
           -t_{1} &             0 &      0 &         t_{2} &             0 &      0 & \Delta &      0 & -t_{1}\\
                0 &        -t_{1} &      0 &             0 &         t_{2} &      0 &      0 & \Delta & -t_{2}\\
                0 &             0 & -t_{1} &             0 &             0 &  t_{2} & -t_{1} & -t_{2} &     0

\end{array}\right)\;. \label{wideeq}
\end{equation}

We calculate again the ground state energy up to $t^{4}$ order, obtaining:

\begin{eqnarray}
E_{\uparrow\downarrow}=-\frac{2t^{2}}{\Delta}+\frac{2t^{4}}{\Delta[(\Delta+\frac{U_{p}}{2})^{2}-\frac{(J_{H}^{Ox})^{2}}{4}]}
-\frac{t^{4}J_{H}^{Ox}sin^{2}(\theta)}{2\Delta^{2}[(\Delta+\frac{U_{p}}{2})^{2}-\frac{(J_{H}^{Ox})^{2}}{4}]}
+\frac{3t^{4}U_{p}}{\Delta^{2}[(\Delta+\frac{U_{p}}{2})^{2}-\frac{(J_{H}^{Ox})^{2}}{4}]}\nonumber\\
-\frac{t^{4}(J_{H}^{Ox})^{2}sin^{2}(\theta)}{4\Delta^{2}(\Delta+\frac{U_{p}}{2})[(\Delta+\frac{U_{p}}{2})^{2}-\frac{(J_{H}^{Ox})^{2}}{4}]}
+\frac{t^{4}[2(U_{P})^{2}-(J_{H}^{Ox})^{2}]}{2\Delta^{3}[(\Delta+\frac{U_{p}}{2})^{2}-\frac{(J_{H}^{Ox})^{2}}{4}]}
-\frac{t^{4}(J_{H}^{Ox})^{2}U_{p}}{4\Delta^{3}(\Delta+\frac{U_{p}}{2})[(\Delta+\frac{U_{p}}{2})^{2}-\frac{(J_{H}^{Ox})^{2}}{4}]}
\end{eqnarray}

The difference of the leading eigenvalues is:

\begin{eqnarray}
J=E_{\uparrow\uparrow}-E_{\uparrow\downarrow}=
\frac{2t^{4}cos^{2}(\theta)}{\Delta[(\Delta+\frac{U_{p}}{2})^{2}-\frac{(J_{H}^{Ox})^{2}}{4}]}
+\frac{t^{4}[U_{p}cos^{2}(\theta)-(\frac{J_{H}^{Ox}}{2})sin^{2}(\theta)]}{\Delta^{2}[(\Delta+\frac{U_{p}}{2})^{2}-\frac{(J_{H}^{Ox})^{2}}{4}]}
-\frac{t^{4}(J_{H}^{Ox})^{2}(2-sin^{2}(\theta))}{4\Delta^{2}(\Delta+\frac{U_{p}}{2})[(\Delta+\frac{U_{p}}{2})^{2}-\frac{(J_{H}^{Ox})^{2}}{4}]}
\end{eqnarray}

\narrowtext
Thus, to the AF result of Ref.\ \onlinecite{mishra98} adds now other terms, which depending on the  $\Delta$, $J_{H}^{Ox}$  and $U_{p}$  values, can yield a ferromagnetic coupling for the bond angle of these pyrochlores, without the necessity of invoking more complicated mechanisms\cite{mishra98}.
As expected $J_{H}^{Ox}$  must have a significant value, as it is the case for other oxides in which this parameter also induces a ferromagnetic interaction, e.g. $NaNiO_{2}$ with $\theta\sim95^{\circ}$ for n.n. $Ni$ ions within the same layer \cite{chappel00}.
For $\theta=135^{\circ}$ the condition is :

\begin{equation}
J_{H}^{Ox}>\frac{4}{3}(\Delta+\frac{U_{p}}{2})
\end{equation}
which means that the $O(2p)$ levels must be close to the $Mn(e_{g})$ state. In Fig. 1 we plot the angular dependence of Eq.(5) for possible parameters for the $A_{2}Mn_{2}O_{7}$
pyrochlores.\\
This first calculation shows that the superexchange alone can explain the ferromagnetic ordering of these compounds at $\sim15K$. Furthermore, the observed decrease of $T_{C}$ with pressure, inducing an isomorphic reduction of the lattice parameters \cite{n\'{u\~{n}ez99}}, appears as a confirmation of this mechanism :
the directional $e_{g}$  levels of the $Mn$ ions will go up in energy, instead the intra-atomic $J_{H}^{Ox}$  will not significantly change, the increase of  the $Mn-O$  hopping term enhancing the resulting
effect. Fig. 2 shows the change of the exchange interaction $J$ with $\Delta$, which is an increasing parameter with the applied pressure.
In real systems the variation will be stronger due to the simultaneous increase of $t$.\\
However, the systematic study performed by Shimakawa et al. \cite{shimakawa99} evidences that there is an additional feature determining $T_{C}$, since the $A=In$, $Tl$ compounds, with similar $Mn-O$ distances and angles to those of  $A=Y$, $Lu$, have both  $T_{C}\sim125K$.
The important difference between both groups of pyrochlores is the nature of the $A$ atom surrounded by inequivalent oxygen ions (we call $O(II)$ the closer ones), which is reflected in the corresponding band
structures.\cite{shimakawa99} For the Y compound the first unoccupied states correspond to the $Mn(e_{g})$ levels, the $Y-O(II)-O(I)$ states having higher energies, i.e. the crystal field splitting is much smaller than the exchange gap. Then the only relevant mechanism is the one calculated before, yielding $T_{C}\sim15K$.\\
In contrast, the $In(5s)-$ and $Tl(6s)-O(II)-O(I)$ bands are lower in energy and strongly hybridised with the empty $e_{g}$ orbital \cite{singh97}. Therefore, the $t_{2g}$ electrons can go to these states of energy $\Delta'$ (we call $t'$ the corresponding matrix element) and due to the large $J_{H}^{Mn}$ this hopping will induce a strong ferromagnetic interaction $J'$ between the $Mn$
ions. We can estimate:

\begin{equation}
J'\simeq\frac{t'^{2}}{\Delta'}\sim110K
\end{equation}
which yields $t'\sim0.15eV$ for $\Delta'\sim2eV$.\cite{shimakawa99} The majority-spin empty band will be strongly polarised parallel to the moment of the $Mn$ ions.\\
On the other hand, we can understand why the conduction band in $Tl_{2}Mn_{2}O_{7}$ is strongly polarised in the opposite direction: it will lower its energy by hybridisation with the $O(I)$ states that do not participate in the superexchange ferromagnetic coupling.\\
In those cases, in which the conduction band approaches (as in A=In) or crosses the Fermi level (for A=Tl), we expect  a relative smaller decrease of the critical temperature (as observed \cite{sushko99}) and even a rise of  $T_{C}$ for higher pressures,
when these bands are modified lowering further their energies.
Recent measurements \cite{n\'{u\~{n}ez99}} confirm this change of tendency for $Tl_{2}Mn_{2}O_{7}$, i.e. they show an initial decrease  followed  by an increase of $T_{C}$ with applied pressure, the minimum critical temperature corresponding to
$\sim10Kbar$. We expect a similar behaviour for A=In, without an enhanced change in the magnetism due to the eventual
insulator-metal transition induced by higher pressure.
The increase of the ferromagnetic coupling in $Tl_{2}(Mn_{2-x}Sb_{x})O_{7}$ (the $Sb$-substitution acting as a negative pressure) can be understood within the same picture, more easily than considering \cite{alonso99} an AF superexchange coupling between n.n. $Mn$ ions.\\
We conclude that when the oxygen intra-atomic interactions are properly included in the calculations, the superexchange between n.n.$Mn$ ions suffices to explain the ferromagnetic $T_{C}'s\sim15K$ of $A_{2}Mn_{2}O_{7}$ pyrochlores, as well as their decrease with applied pressure. Instead, an additional mechanism
involving the $In(5s)/Tl(6s)-O(2p)-Mn(e_{g})$ bands and the $t_{2g}$ electrons is neccessary to account for the higher
$T_{C}'s\sim125K$ and their non-monotonic behaviour with pressure.

\narrowtext

\narrowtext

\begin{figure}
\caption{Angle dependence of the n.n. $Mn-O(I)-Mn$ exchange interaction $J$, Eq.5, for typical
parameters: $\Delta=0.2eV$, $J_{H}^{Ox}=3eV$, $U_{p}=4eV$ and $t=0.1eV$.}
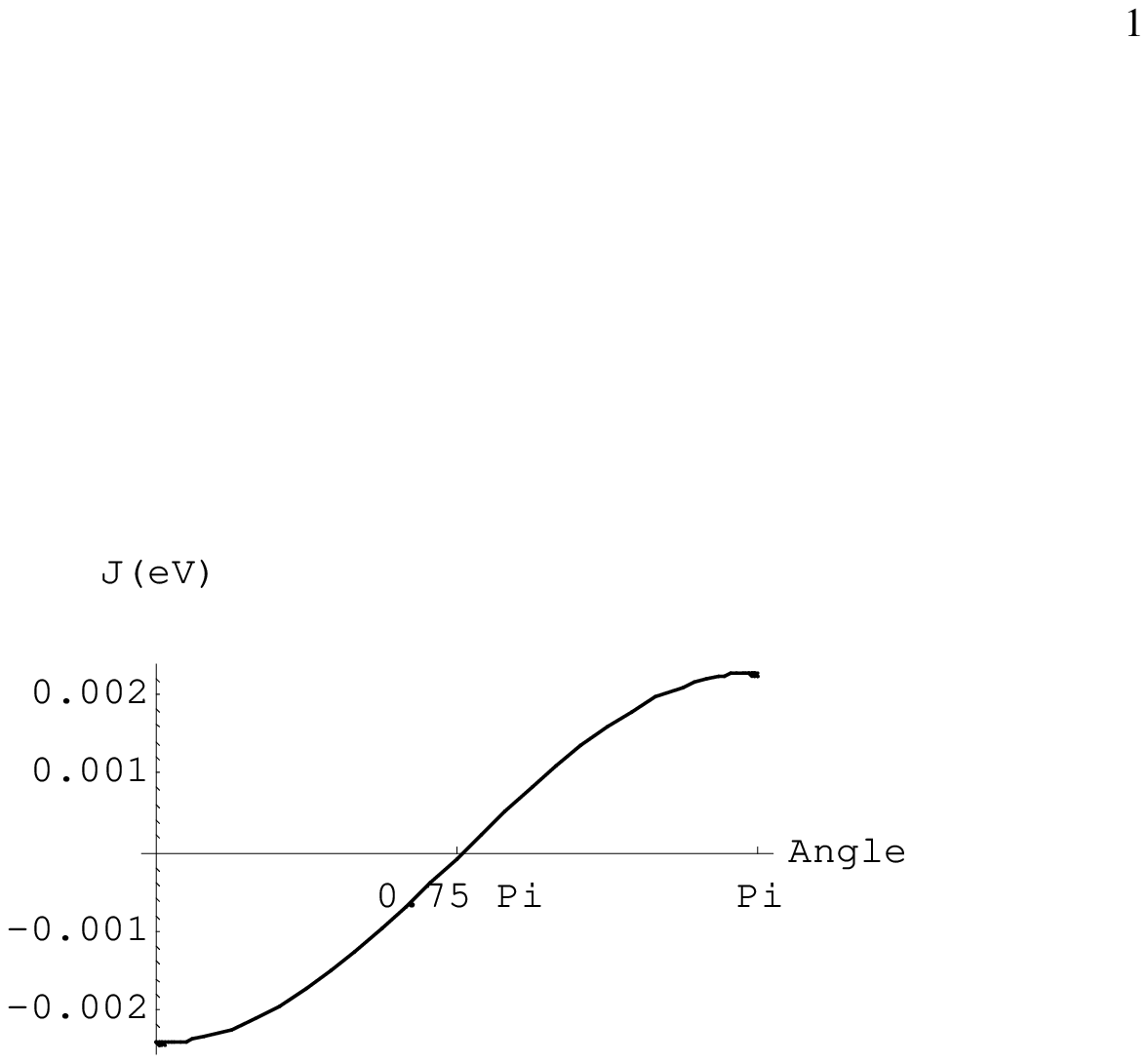
\label{A2Mn2O7fig1.eps}
\end{figure}

\begin{figure}
\caption{Variation of $J$ with the energy difference $\Delta$ between $O(2p)$ levels and the $Mn(e_{g})$ states, given by Eq.5,
for $J_{H}^{Ox}=3eV$, $U_{p}=4eV$, $t=0.13eV$ and $\theta=133^{\circ}$.}
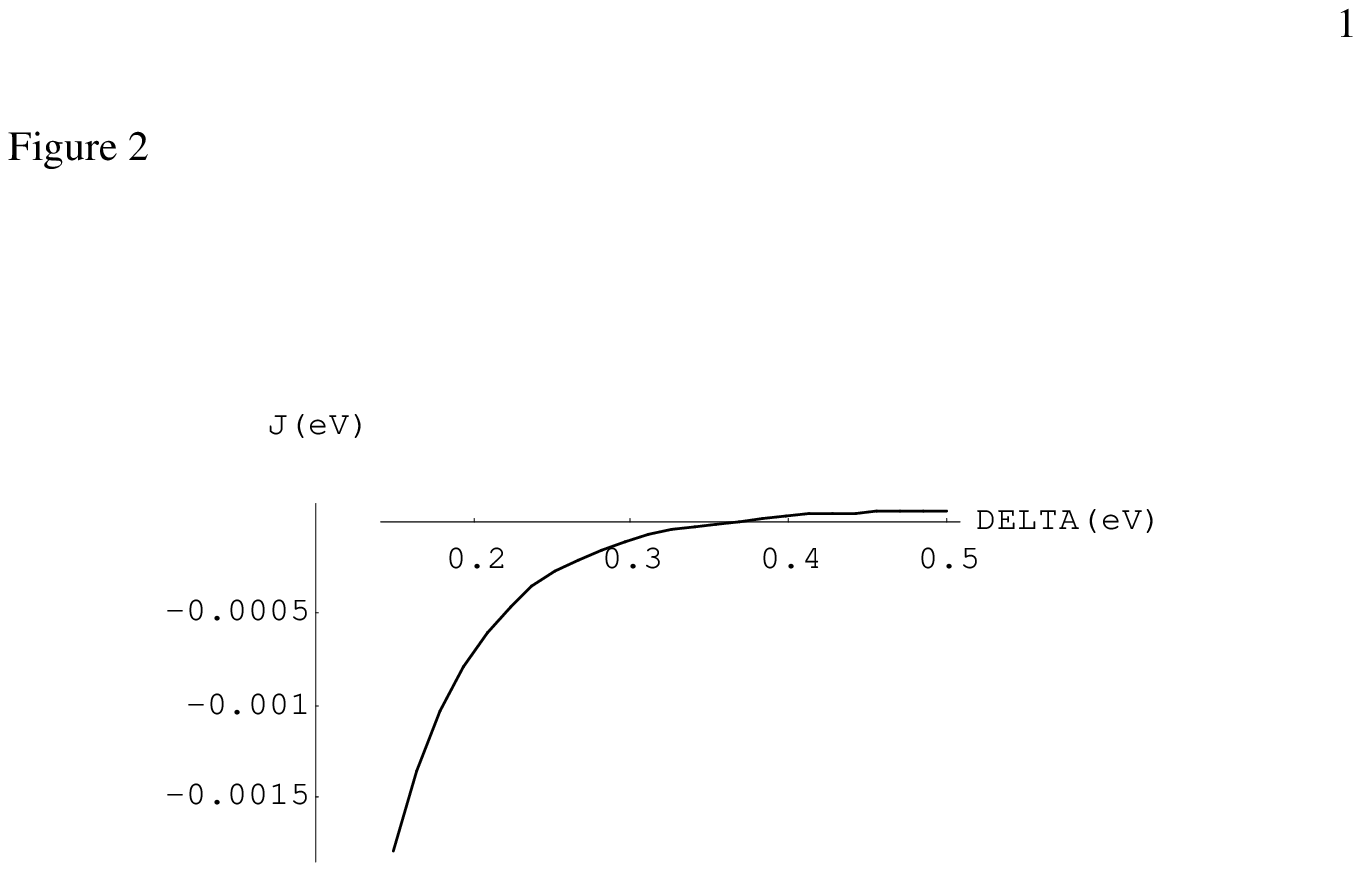
\label{ A2Mn2O7fig2.eps}
\end{figure}

\end{document}